\documentclass[]{spie}  

\usepackage{amsmath,amsfonts,amssymb}
\usepackage{graphicx}
\usepackage[colorlinks=true, allcolors=blue]{hyperref}

\title{Progress on the SOXS transients chaser for the ESO-NTT}

\author[a]{P.~Schipani}
\author[b]{S.~Campana}
\author[c]{R.~Claudi}
\author[b]{M.~Aliverti}
\author[a]{A.~Baruffolo}
\author[d]{S.~Ben-Ami}
\author[a]{G.~Capasso}
\author[e]{R.~Cosentino}
\author[f]{F.~D'Alessio}
\author[b]{P.~D'Avanzo}
\author[d]{O.~Hershko}
\author[g,h]{H.~Kuncarayakti}
\author[b]{M.~Landoni}
\author[k]{M.~Munari}
\author[m,n]{G.~Pignata}
\author[c]{K.~Radhakrishnan}
\author[o,d]{A.~Rubin}
\author[p]{S.~Scuderi}
\author[f]{F.~Vitali}
\author[q]{D.~Young}
\author[r]{J.~Achrén}
\author[s]{J.~A.~Araiza-Duran}
\author[t]{I.~Arcavi}
\author[c]{F.~Battaini}
\author[s]{A.~Brucalassi}
\author[d]{R.~Bruch}
\author[c]{E.~Cappellaro}
\author[a]{M.~Colapietro}
\author[a]{M.~Della~Valle}
\author[k]{R.~Di~Benedetto}
\author[a]{S.~D'Orsi}
\author[d]{A.~Gal-Yam}
\author[b]{M.~Genoni}
\author[e]{M.~Hernandez}
\author[h,g]{J.~Kotilainen}
\author[u]{G.~Li~Causi}
\author[c]{L.~Lessio}
\author[a]{L.~Marty}
\author[g]{S.~Mattila}
\author[d]{M.~Rappaport}
\author[c]{D.~Ricci}
\author[b]{M.~Riva}
\author[c]{B.~Salasnich}
\author[a]{S.~Savarese}
\author[q]{S.~Smartt}
\author[k]{R.~Zanmar~Sanchez}
\author[v]{M.~Stritzinger}
\author[z]{G.~Umbriaco}
\author[e]{H.~Ventura}
\author[o]{L.~Pasquini}
\author[o]{M.~Sch\"oller}
\author[o]{H.-U.~K\"aufl}
\author[o]{M.~Accardo}
\author[o]{L.~Mehrgan}
\author[o]{E.~Pompei}
\author[o]{I.~Saviane}

\affil[a]{INAF -- Osservatorio Astronomico di Capodimonte, Sal. Moiariello 16, I-80131, Naples, Italy }
\affil[b]{INAF -- Osservatorio Astronomico di Brera, Via Bianchi 46, I-23807, Merate, Italy }
\affil[c]{INAF -- Osservatorio Astronomico di Padova, Vicolo dell’Osservatorio 5, I-35122, Padua, Italy }
\affil[d]{Weizmann Institute of Science, Herzl St 234, Rehovot, 7610001, Israel }
\affil[e]{FGG-INAF, TNG, Rambla J.A. Fernández Pérez 7, E-38712 Breña Baja (TF), Spain }
\affil[f]{INAF -- Osservatorio Astronomico di Roma, Via Frascati 33, I-00078 M. Porzio Catone, Italy }
\affil[g]{Tuorla Observatory, Dept. of Physics and Astronomy, University of Turku,  FI-20014, Finland }
\affil[h]{Finnish Centre for Astronomy with ESO (FINCA), FI-20014 University of Turku, Finland }
\affil[k]{INAF -- Osservatorio Astrofisico di Catania, Via S. Sofia 78, I-95123 Catania, Italy }
\affil[m]{Universidad Andres Bello, Avda. Republica 252, Santiago, Chile }
\affil[n]{MAS, Nuncio Monseñor Sotero Sanz 100, Providencia, Santiago, Chile }
\affil[o]{ESO, Karl Schwarzschild Strasse 2, D-85748, Garching bei München, Germany }
\affil[p]{INAF - Istituto di Astrofisica Spaziale e Fisica Cosmica, Via Corti 12, I-20133, Milano, Italy}
\affil[q]{Astrophysics Research Centre, Queen's University Belfast, Belfast, BT7 1NN, UK }
\affil[r]{Incident Angle Oy, Capsiankatu 4 A 29, FI-20320 Turku, Finland }
\affil[s]{INAF - Osservatorio Astrofisico di Arcetri, Largo E. Fermi 5, I-50125, Firenze, Italy}
\affil[t]{Tel Aviv University, Department of Astrophysics, 69978 Tel Aviv, Israel }
\affil[u]{INAF - Istituto di Astrofisica e Planetologia Spaziali, I-00133 Rome, Italy}
\affil[v]{Aarhus University, Ny Munkegade 120, D-8000 Aarhus, Denmark}
\affil[z]{Univ. Padova, Dept. Physics and Astronomy, Vicolo dell'Osservatorio 3, I-35122 Padua, Italy}

\authorinfo{Send correspondence to: pietro.schipani@inaf.it}

\begin{document} 
\maketitle

\begin{abstract}
SOXS (Son Of X-Shooter) is a single object spectrograph offering a simultaneous spectral coverage from U- to H-band, built by an international consortium for the 3.58-m ESO New Technology Telescope at the La Silla Observatory. It is designed to observe all kind of transients and variable sources discovered by different surveys with a highly flexible schedule maintained by the consortium, based on the Target of Opportunity concept. SOXS is going to be a fundamental spectroscopic partner for any kind of imaging survey, becoming one of the premier transient follow-up instruments in the Southern hemisphere. This paper gives an updated status of the project, when the instrument is in the advanced phase of integration and testing in Europe, prior to the activities in Chile.
\end{abstract}


\begin{figure} [ht]
\begin{center}
\begin{tabular}{c} 
\includegraphics[height=11.9cm]{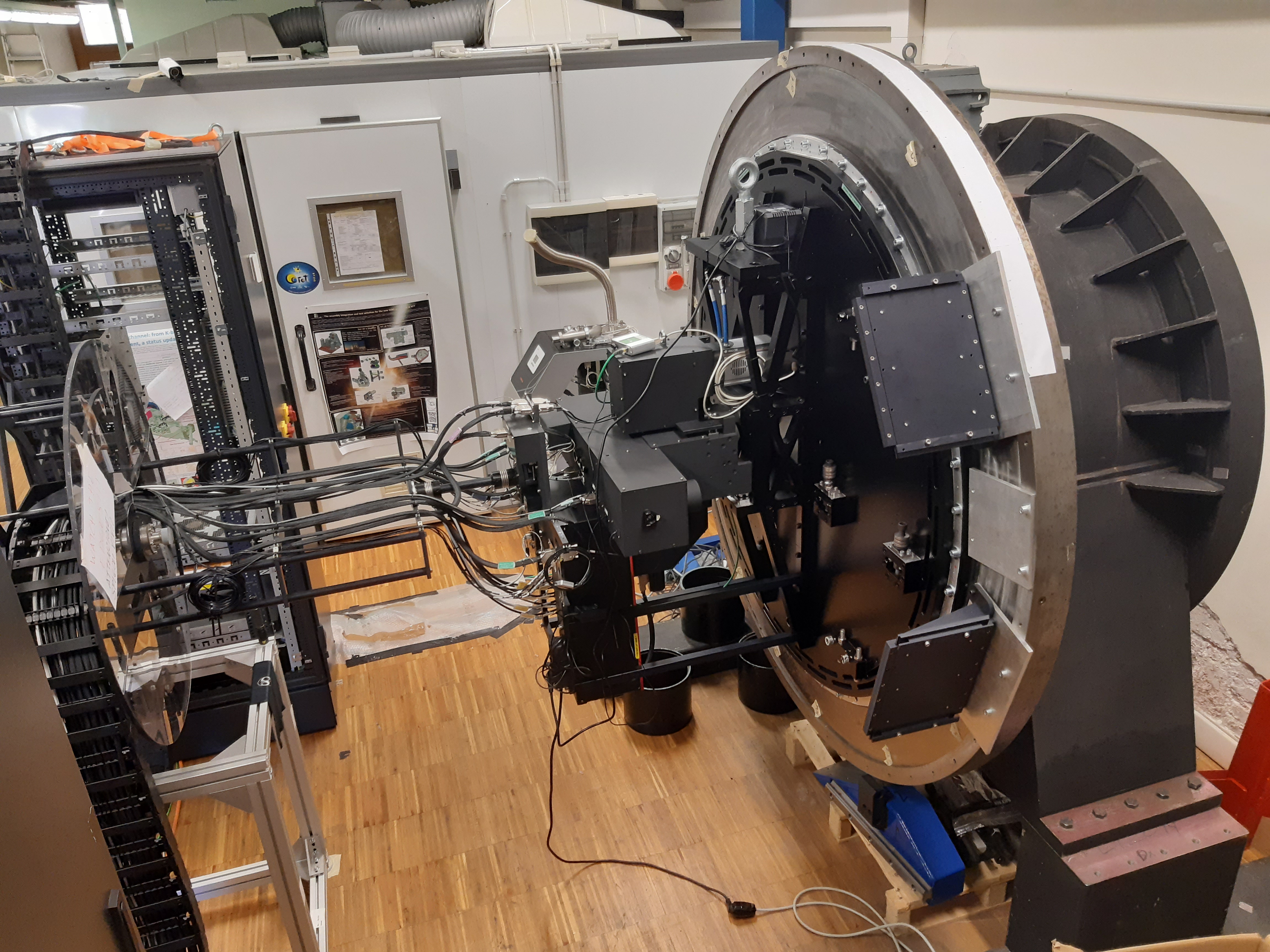}
\end{tabular}
\end{center}
\caption[example1] 
{ \label{fig:SOXSRight} 
The SOXS integration room with the instrument partially assembled. The NTT simulator provides the same interfaces of the real telescope. The instrument can be rotated and tested in all gravity conditions. }
\end{figure} 

\begin{figure} [ht]
\begin{center}
\begin{tabular}{c} 
\includegraphics[height=10cm]{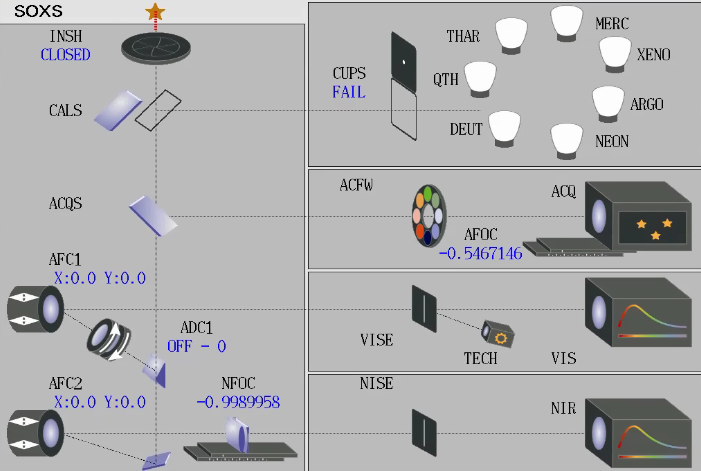}
\end{tabular}
\end{center}
\caption[example1] 
{ \label{fig:SOXSSynopt} 
The control software synoptic diagram.}
\end{figure} 

\begin{figure} [ht]
\begin{center}
\begin{tabular}{c} 
\includegraphics[height=10cm]{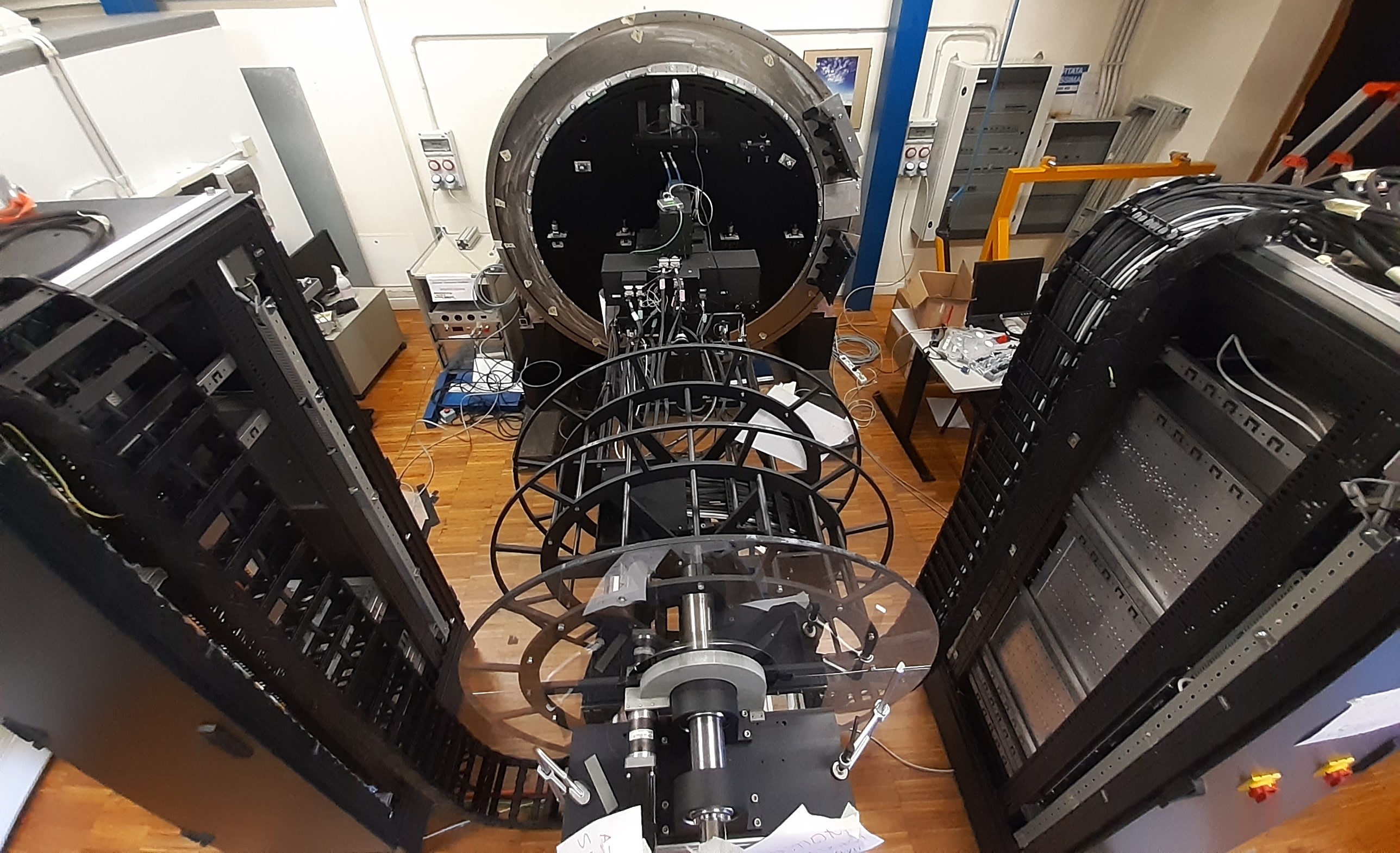}
\end{tabular}
\end{center}
\caption[example1] 
{ \label{fig:Corot} 
The corotator system includes two cable-chains routing the cables from the two electronic cabinets to the instrument.}
\end{figure}

\section{INTRODUCTION}
\label{sec:intro}  
The SOXS instrument has been presented in previous contributions, reporting on the status of the project at regular intervals\cite{Schipani16,Schipani18,Schipani20}. Thus, the instrument design is only briefly sketched in this article, where we focus more on the progress of the SOXS project, that is now in the Assembly, Integration and Test phase in Italy (Fig.~\ref{fig:SOXSRight}).

The SOXS consortium includes institutes from 6 countries in 3 continents, Italy, Israel, UK, Chile, Finland and Denmark, with INAF (IT) acting as leading institute. The consortium will receive 180 nights/year of observing time in return of the realization of the instrument. SOXS  will be offered to the ESO community for the remaining time. All observation proposals will go through the regular ESO approval process. The consortium will manage the operations of SOXS remotely, through dedicated tools developed in parallel with the instrument hardware. 

The NTT-SOXS combination is one of the two pillars of the ESO strategy for the future of the historic La Silla observatory, based on the specialization of the two 4-m class telescopes to well-defined science cases. The NTT is dedicated to transient sources and the ESO 3.6-m telescope to extrasolar planets.

The consortium will use SOXS for the classification and characterization of all kind of astrophysical transients, e.g. supernovae, electromagnetic counterparts of gravitational wave events, neutrino events, tidal disruptions of stars in the gravitational field of supermassive black holes, gamma-ray and fast radio bursts, X-ray binaries and novae, magnetars, but also asteroids and comets, activity in young stellar objects, blazars, and AGN.

This science case is well served by a single-object, wide-band and medium resolution spectrograph. SOXS is designed with an average  R$\sim$4500 for a 1 arcsec slit, capable of simultaneously observing over the spectral range 0.35-2.0$ \mu$m. The spectroscopic mode of SOXS is based on the simultaneous observation of the same target with two distinct spectrographs, 
one working in the UltraViolet-Visible (350-850 nm) and the other in the Near InfraRed (800-2000 nm) bands. The wavelengths overlap by design to allow for the cross-calibration of the two arms. The UV-VIS spectrograph is based on an innovative multi-grating concept\cite{Rubin18}, imaging four different narrow band spectra on a single  camera\cite{Oliva16}, whereas the NIR arm\cite{Sanchez18} implements a 4C layout with Collimator Compensation of Camera Chromatism\cite{Delabre89}, where the  main dispersion is implemented by a reflection grating and prisms in double-pass provide the cross-dispersion, imaging a classic cross-dispersed echellogramme.

The two spectrographs are fed by a common opto-mechanical system, the ``Common Path", which receives the light at the telescope focus and splits it to the two spectrograph slits, through relay optics which reduce the F/number from F/11 to F/6.5. The atmospheric dispersion is corrected, but only in the UV-VIS arm where the effect is more significant. Additionally, the Common Path drives the light to the $3.5'\times 3.5'$ acquisition and guiding camera that, indeed, is also a scientific camera that implements the imaging mode of SOXS.
The unit for the wavelength and flux calibration is also connected to the Common Path structure, featuring a set of light sources covering all possible needs in the wide wavelength range.
The functional diagram of the instrument is represented in Fig.~\ref{fig:SOXSSynopt}, which is a control software panel providing all the relevant status information.
More details on each subsystem can be found in a set of related articles\cite{Aliverti18,Biondi18,Brucalassi18,Capasso18,Claudi18,Cosentino18,Ricci18,Rubin18,Sanchez18,Vitali18,Aliverti20,Biondi20,Brucalassi20,Claudi20,Colapietro20,Cosentino20,Genoni20,Kuncarayakti20,Ricci20,Rubin20,Sanchez20,Vitali20,Young20}.

\section{Manufacturing}
\label{sec:manufacturing} 
The modular design of SOXS allowed for a parallel development in different countries of its subsystems:
\begin{itemize}
    \item UV-VIS spectrograph (Israel)
    \item NIR spectrograph (Italy)
    \item Common Path (Italy)
    \item Acquisition and Guiding Camera (Chile)
    \item Calibration Unit (Finland)
\end{itemize}

Other common parts, e.g. the control electronics, the support structure and the corotator, were built in Italy. Notably, the parts built in Italy were all provided by INAF but through six different institutes. 
The procurement were even more geographically distributed.
In fact, all partners were independent in the choice of suppliers and the budget size of the project made big subcontracting not sufficiently attractive for suppliers. Thus, the SOXS partners have worked with a number of different suppliers; e.g. the production of the optical elements involved about 10 different companies and the consortium partners performed the system integration. Contracts were signed with suppliers based in the consortium countries, but also from e.g. US, Japan, Mexico, Germany. Nothing strange in normal times when traveling is possible, but the start of pandemic not only caused delays in all deliveries, but also prevented any review in person of the parts under development at suppliers premises. Looking retrospectively, maybe this latter issue was the worst, although there was no way to predict and mitigate this risk.

\begin{figure} [ht]
\begin{center}
\begin{tabular}{c} 
\includegraphics[height=10cm]{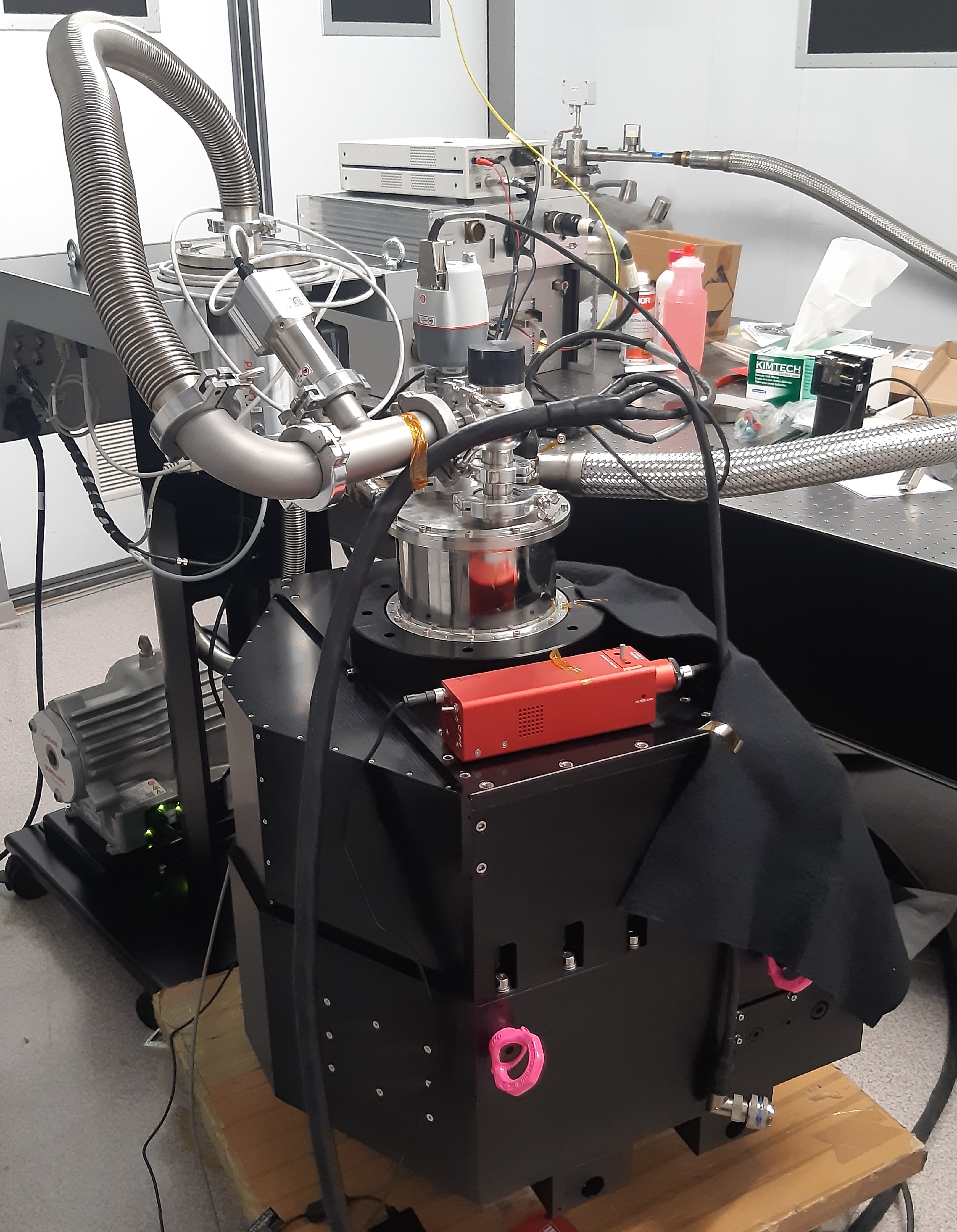}
\includegraphics[height=10cm]{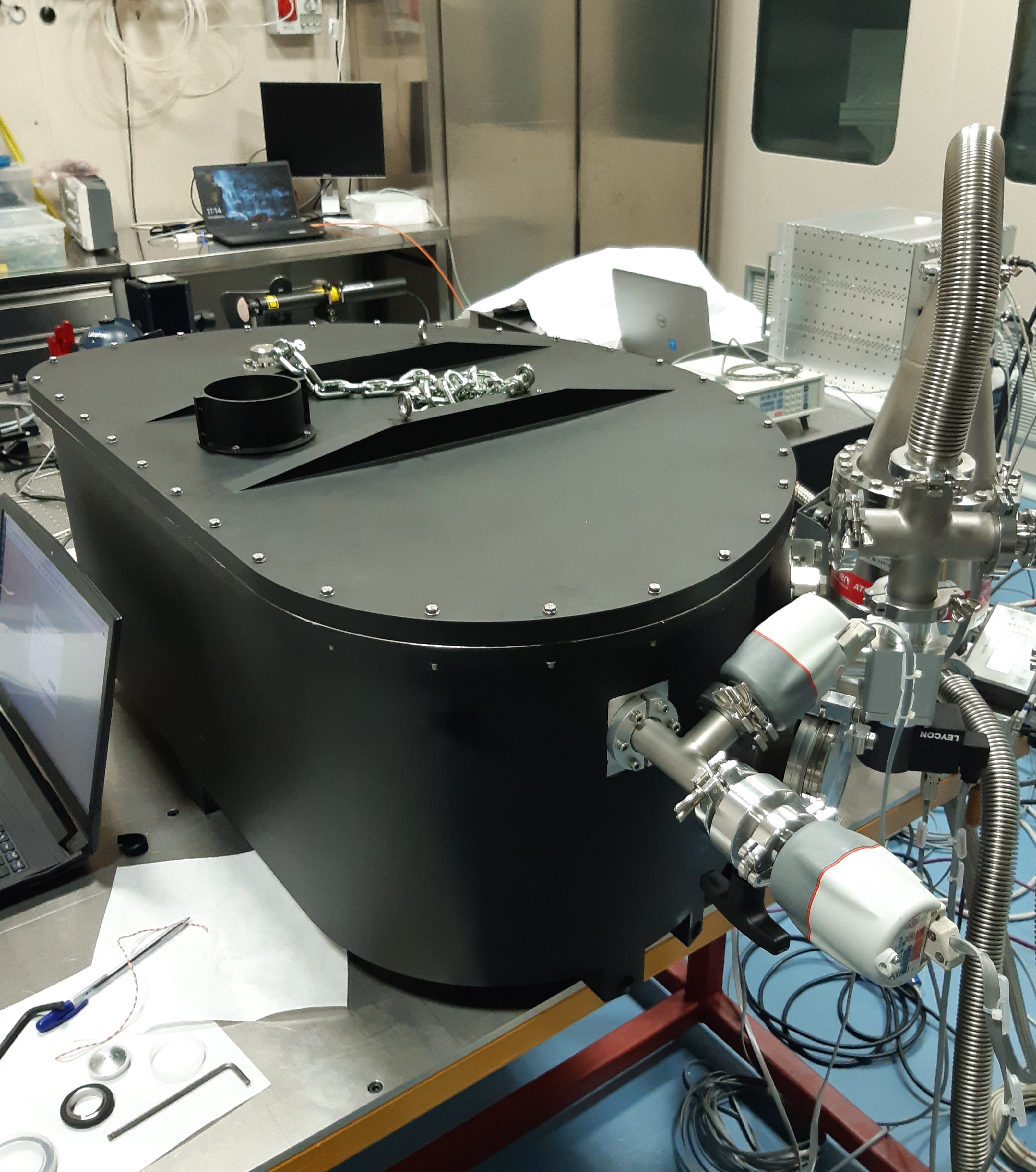}
\end{tabular}
\end{center}
\caption[example1] 
{ \label{fig:Spectrographs} 
The UV-VIS (left) and NIR (right) spectrographs under tests.}
\end{figure} 

\begin{figure} [ht]
\begin{center}
\begin{tabular}{c} 
\includegraphics[height=11.9cm]{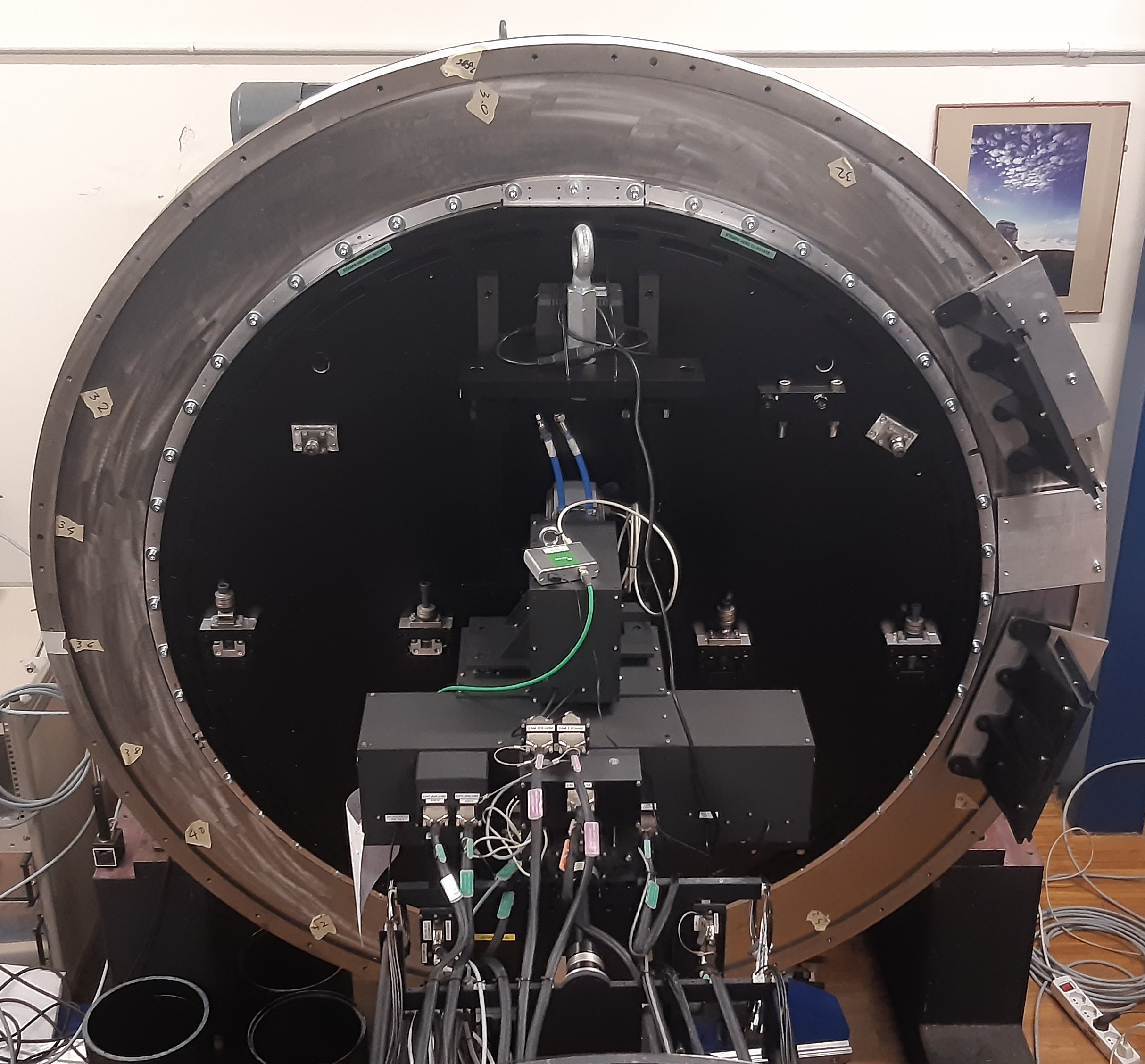}
\end{tabular}
\end{center}
\caption[example1] 
{ \label{fig:SOXSFront} 
The Common Path, Acquisition Camera and Calibration Unit installed on the instrument flange.}
\end{figure} 

\section{Status}
\label{sec:status} 

\subsection{The structure}
The interface flange is completed and mounted on the NTT simulator. The SOXS platform will be installed in Chile, replacing the existing Nasmyth platform at the NTT. No installation of the platform is foreseen in Italy, given the integration area capabilities. The corotator mechanics and control system is mounted, tested, and fully operational (Fig.~\ref{fig:Corot}).

\subsection{Common Path}
The Common Path has been aligned with satisfactory results and is fully available. An issue with the Atmospheric Dispersion Corrector was solved after a long process involving disassembly, realignment and new assembly of the two single counter-rotating quadruplets which compose the system, followed by alignment and tests of the full combined system.

\subsection{UV-VIS Spectrograph}
The UV-VIS spectrograph has been realized at the Weizmann institute and then moved to Italy for the integration of the detector and the final tests of the whole subsystem. Fig.~\ref{fig:Spectrographs} shows it right after its arrival to Italy.

\subsection{NIR Spectrograph}
The NIR spectrograph is under tests at INAF-Merate. It is shown in Fig.~\ref{fig:Spectrographs}, as well. After a long phase of procurement all the parts are available and the current work is about the development of the cryo-vacuum control procedures. The very next phase includes the alignment of the optics and the installation and test of the H2RG detector.

\subsection{Acquisition Camera}
The Acquisition and Guiding Camera is available at the integration site. It is visible in Fig.~\ref{fig:SOXSFront} on top of the Common Path. All the functionalities have been tested and the optics alignment is proceeding.

\subsection{Calibration Unit}
The Calibration Unit has been delivered to Italy and functionally tested, so it is available at the integration site (see Fig.~\ref{fig:SOXSFront} on the bottom side of the Common Path). Some work is ongoing to improve its accessibility before going to Chile.

\subsection{Instrument Software}
The Instrument Software has been developed using the VLT Common Software framework. In parallel with the increasing availability of the hardware, the software has evolved making possible to operate and test the subsystems. Some templates have been tested with a consistent part of the hardware available, as most of the motorizations are inside the Common Path, that is completed.

\begin{figure} [ht]
\begin{center}
\begin{tabular}{c} 
\includegraphics[height=8.6cm]{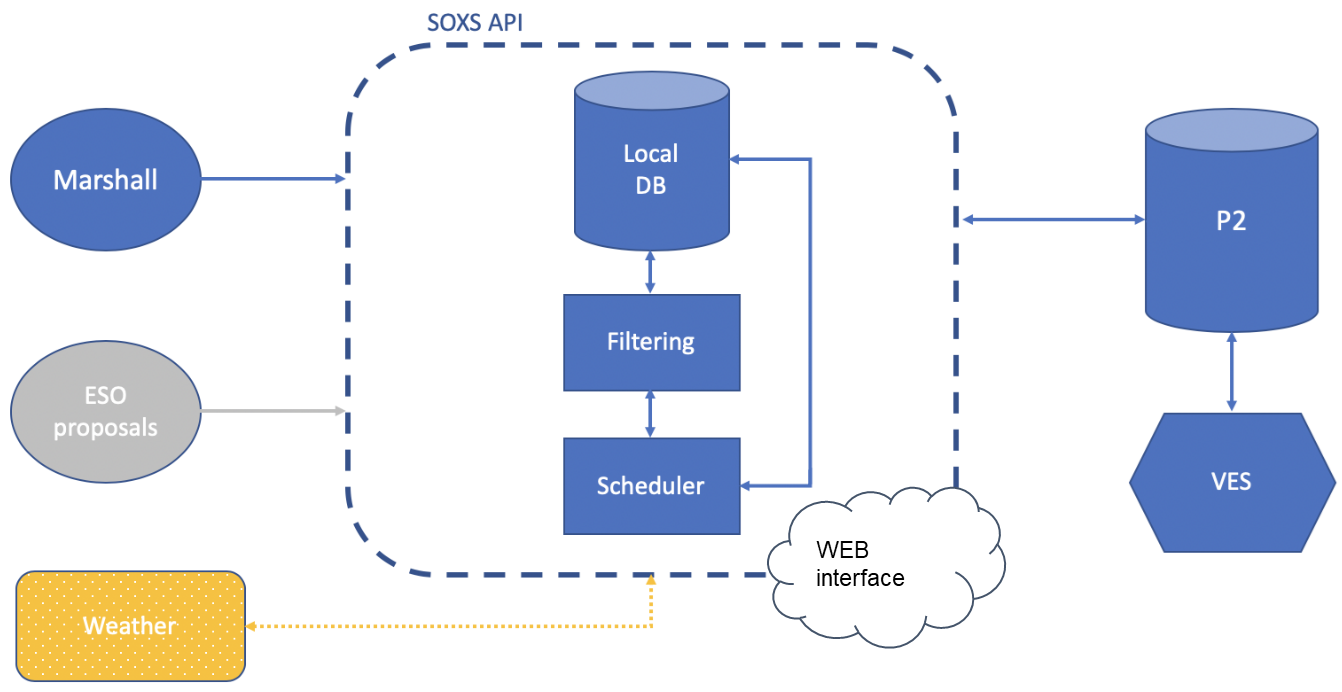}
\end{tabular}
\end{center}
\caption[example1] 
{ \label{fig:Scheduler} 
Architecture of the SOXS Scheduler.}
\end{figure} 

\subsection{Control Electronics}
The instrument control system is modular and composed by several 19" subracks, hosted within two cabinets placed on opposite sides (Fig.~\ref{fig:Corot}). Only minimal electronic parts are placed on the instrument, following the guidelines to maintain it as light as possible and remove heat sources. Thus, only the NGC controllers provided by ESO are placed close to the detectors to minimize the cable lengths. 

A Beckhoff PLC drives all the instrument functions. It is hosted in one of the two cabinets and is under continuous work during the ongoing tests of the instrument. It follows the ESO electronic standards for the new instruments. The cabling has been routed through the cable chains and the combined rotation of the instrument and the corotator is fully operational.

The cryo-vacuum is a fully independent system. A Siemens PLC is the heart of the cryo-vacuum control electronics. At the time of writing, it is under intensive tests serving the NIR spectrograph, that is the last subsystem waiting to be integrated with the others. Therefore, a temporary backup cryo-vacuum control system has been provided to work in parallel with the rest of the electronics visible in Fig.~\ref{fig:Corot}, serving the UV-VIS detector cryostat.

\section{Tools for the operation phase}
\label{sec:opera} 
The consortium is in charge of the operations of the instrument. Thus, many activities are in progress to develop the tools to adopt in that phase, in parallel to the physical realization of the spectrograph.

\subsection{Scheduler}
The consortium will manage the night schedule of the telescope. Given the focus on the transient sources, the schedule will be highly dynamic with a permanent use of the Target of Opportunity concept. A telescope operator on-site will execute the Observation Blocks proposed by a scheduler. The scheduler is a web application receiving inputs from three main sources:
\begin{itemize}
    \item Marshall
    \item ESO community proposals
    \item Astronomical Site Monitor
\end{itemize}

The Marshall is an evolution of the application developed within the PESSTO\cite{Smartt13} project, including information about transients discovered by a number of all-sky surveys. In the next future, it is expected the Rubin telescope will provide a huge amount of possible targets for SOXS. However, there are several very good feeders for this job (ZTF, ATLAS, etc.) even before the Rubin era. The most interesting targets are selected, transformed to Observation Blocks, merged with OBs provided by the ESO community. The scheduler reads continuously the weather and seeing information from the local Astronomical Site Monitor. This makes dynamic changes to the schedule possible. The scheduler is under development and basic interface tests with Chile have been completed successfully. 
The schedule will be completely dynamical and can be changed on the fly, if an interesting transient sets in.

\subsection{ETC}
The consortium has developed the Exposure Time Calculator, that will be publicly available to all the instrument users, including the ESO community astronomers submitting regular proposals.
The tool has been continuously improved over time by including as-built data for all the relevant instrument components, i.e. optical elements and detectors. It is a ongoing effort which will be  validated on sky. The consortium will maintain the ETC over the duration of the operations contract.

\subsection{E2E Modeling}
An instrument model has been developed and frequently upgraded. It is able to produce synthetic spectra that are used by the pipeline developers to test their recipes.

\subsection{Pipeline}
The pipeline will run automatically on the summit, aiming at delivering reduced data through the ESO archive.
The consortium made a big progress in the pipeline, written in Python. The first recipes have been tested and a first release is going to be ready before the last review in Europe. 

\subsection{Quality Control}
The pipeline automatically calculates quality checks on calibration products, useful to check the current status of the instrument (e.g. RON, dark, spectral format). A daemon running on the pipeline machine will push data into a database which will feed the SOXS Quality Control web page. The basic structure is ready to be debugged with real data.

\subsection{Instrument Monitoring}
All the ESO instruments produce log files in text format, including a gold mine of information on the instruments status. Recently, the usability of such data has been largely improved by the development of Datalab\cite{Pena18}, which allows for an easy graphical display of selected information, once they are written by the instrument software into the log files. The SOXS consortium will use this mechanism to monitor remotely all the relevant parameters, which for the moment have been included into the instrument logs. The tool will be operative once the instrument will be onsite.

\section{Conclusions}
\label{sec:concl}
SOXS is going to work at the ESO La Silla observatory as a precious facility for the spectroscopic follow-up of transient sources. The last years of pandemic caused delays to the project, which had just started the procurement and the integration of the subsystems in the SOXS laboratories. All planned team activities in presence in the laboratories were greatly delayed during the realization phase of the subsystems.

However, the preliminary integration in Italy is now in an advanced status. It will be followed by a period of system tests, the ESO Preliminary Acceptance in Europe process and  the shipment to Chile planned in 2023. 

\bibliography{SOXS} 
\bibliographystyle{spiebib} 

\end{document}